\documentclass{article}  
\usepackage{jamaica04}
\frompage{000} \topage{000}                                              
\def\rightmark{Freeze-out dynamics at RHIC}
\def\leftmark{O. Barannikova}

\def \dndy  {dN/dy}

\def \mbeta {\langle \beta \rangle}
\def \mEt   {\langle E_{T} \rangle}

\title{Freeze-out dynamics at RHIC} 
\authors{
{Olga Barannikova$^1$ for the STAR Collaboration %
}\\[2.812mm]
{\normalsize
 \hspace*{-8pt}$^1$ Purdue University, Department of Physics \\ 
West Lafayette, IN 47906, USA\\[0.2ex] 
}}
 
\abstract{
Investigation of the final hadronic state properties of ultra-relativistics pp and Au+Au collisions supplies information on freeze-out conditions at RHIC and possible insights into early stages of these collisions. A variety of particle spectra measured by STAR are studied within the framework of chemical and local kinetic equlibrium models. Here we present the extracted chemical and final kinetic freeze-out temperatures, strangeness saturation factor, final collective flow velocity, and the inferred flow velocity at chemical freeze-out. In light of those measurements we  disscuss dynamical evoution of the collission system.
}

 
\begin{document}
 
\maketitle
\setcounter{page}{1}

\section{Introduction}\label{intro}

Existence of the new state of matter - Quark Gluon Plasma (QGP) - has been predicted by Quantum Chromodynamics (QCD) long time ago. Many theoretical works were devoted to understanding of QGP properties, its possible signatures, and relevant phase transition conditions.  Recent  calculations of lattice QCD indicate a value of $T_c \approx 170$~MeV\cite{karsch} for the QGP phase transition critical temperature. Ultra-relativistic heavy ion collisions provide an experimental opportunity to probe nuclear matter under conditions which are believed to be sufficient for  QGP formation. 

The evolution of ultra-relativistic  heavy-ion collisions at RHIC  is generally thought to  proceed  through the following stages: 
$1)$~deconfinment of quark and gluons at high energy density in the initial stage, followed by equilibration of quarks and gluons; $2)$~phase transition and hadronization;  $3)$~hadronic interactions and chemical freeze-out; $4)$~ elastic scattering and thermal (kinetic) freeze-out. 
The bulk of the created medium appears in soft hadrons, which  we study  in this paper in attempt to gain understanding about the dynamics of heavy ion collisions. Identified particle distributions provide valuable information about freeze-out properties of such collisions. They also may contain important traces from the early, possibly partonic stage of the collisions. For example, the final collective transverse radial flow, due to its cumulative nature, contains any collective flow generated at partonic stages, and particle  ratios 
may reflect statistical nature of the hadronization process.  Study of spectral shapes allows to extract system temperature at thermal freeze-out; flavor composition reveals information about strangeness and baryon production and allows to measure the chemical freeze-out temperature and chemical potentials.

    STAR experiment has measured  a variety of hadron species ($\pi^{\pm}$,  $K^{\pm}$, $K^0_s$, $K^*$, $\phi$, $p$, $\bar{p}$, $\Lambda$, $\bar{\Lambda}$,  $\Xi$, $\bar{\Xi}$, $\Omega+\bar{\Omega}$) in pp and Au+Au collisions at 200 GeV~\cite{s1,s2,s3,PRL}. 
The rich set of data provides an unique opportunity to investigate the freeze-out properties and dynamics of heavy ion collisions.

\section{Data Analysis}\label{techno}  

Charged particles are detected in the STAR Time Projection Chamber (TPC)~\cite{starNIM}. TPC is surrounded by a solenoidal magnet, which provided  a uniform magnetic field of 0.5T. Zero degree calorimeters
and beam-beam counters~\cite{highpt} provide a minimum bias trigger for Au+Au and $pp$ collisions, respectively. The Au+Au events are divided into  9  centrality classes based on measured charged particle multiplicity within pseudo-rapidity  $| \eta | < 0.5$. These classes consist, from central to peripheral, of 0-5\%, 5-10\%, 10-20\%, 20-30\%, 30-40\%, 40-50\%, 50-60\%, 60-70\%, and 70-80\% of the geometrical cross-section. For rear particles some of the neighboring centrality bins were combined to improve statistics. 

Particle identification is accomplished by measuring the ionization energy loss $dE/dx$ ($\pi^{\pm}$,  $K^{\pm}$, $p$, $\bar{p}$), decay topology together with invariant mass reconstruction ($K^0_s$,  $\Lambda$, $\bar{\Lambda}$, $\Xi$, $\bar{\Xi}$,  $\Omega+\bar{\Omega}$), and combinatoric invariant mass reconstruction ($\phi$, $K^*$).  Corrections were applied to account for  tracking inefficiency, detector acceptance, hadronic interactions, and particle decays.  Those corrections were obtained from embedding Monte Carlo (MC) tracks into real events at the raw data level and subsequently reconstructing these events.  The propagation of single tracks is calculated using  GEANT, a detailed  description of the STAR geometry, and a realistic simulation of the TPC response \cite{pbar,p,pi_star}. Only pion spectra were further corrected for weak decay products, while the spectra of other particles represent inclusive measurements. 

\section{Results}
\subsection{Kinetic freeze-out properties}\label{details}

At RHIC energies low transverse momentum spectra show clear deviation from the widely used $m_T$-exponential form. Mass-dependent hardening of particle spectra, characteristic of collective flow,  was observed and found to increase  with collision centrality. We use blast-wave model~\cite{BW} in order to describe the spectra and extrapolate our data outside the measured  momentum range. The model provides hydrodynamically motivated description of the phase-space density at kinetic freeze-out, assuming boost-invariant longitudinal flow.
It has been successfully used to describe $p_T$-spectra with three parameters - kinetic freeze-out temperature $T_{kin}$, radial flow velocity $\mbeta$, and the exponent in the flow velocity profile $(\frac{r}{R})^n$.

As shown in Figure 1 (black solid lines) this parameterization is able to simultaneously describe spectra of common particle (pions, kaons and (anti)protons) with a single set of freeze-out parameters at each centrality. Centrality dependence of the extracted fit results is shown in Figure 2: the  more central the collision, the lower extracted temperature and the higher collective radial flow velocity, reaching the values of  $T_{kin}= 89 \pm 10$~MeV and $\langle \beta \rangle = 0.59 \pm 0.05$ in the 5\% most central collisions \cite{PRL}.
 
\begin{figure}[htb]
\insertplot{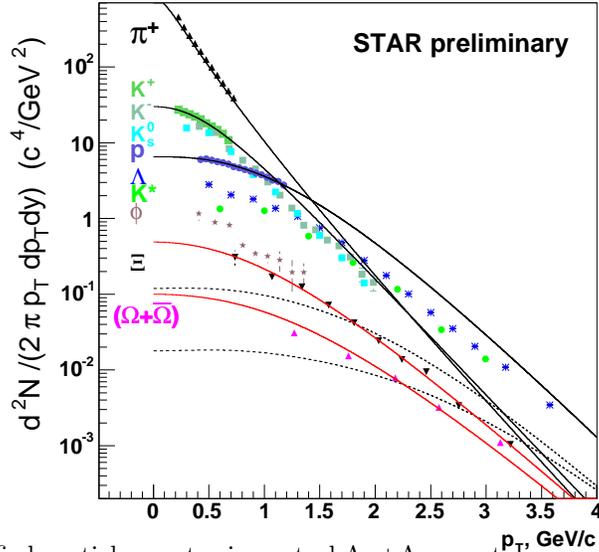}
\vspace*{-1.cm}
\caption[]{Identified particle spectra in  central Au+Au events measured by the STAR experiment. Black solid lines show blast-wave fit results for charged pions, kaons and protons. Dashed lines represent blast-wave prediction for multi-strange baryon spectra shapes based on freeze-out parameters of $\pi$, $K$, $p$. Red lines are blast-wave fits to the $\Xi$ and $\Omega$ spectra.
}
\label{fig1}
\end{figure}

Blast-wave model can also reproduce spectral shapes for other particles measured by STAR, 
but it appears  that different set of freeze-out parameters is needed to describe rare particle distributions. Dashed lines shown in Figure 1 illustrate that blast-wave model fit to the common particle spectra fails to reproduce spectra shape of the rare particles. Red lines show the blast-wave model fir to the $\Xi$ and $\Omega$ spectra, yielding a set of parameters different from that of the common particles. 
 The fit parameters  of such single-spectra fits are shown in Figure 2 in colored symbols. Possible correlation of the lower values of radial flow velocity for rare particles with their low hadronic interaction cross-sections suggests sequential freeze-out of particle species, with multi-strange baryons decoupling from the system earlier than the  common particles, that are more prone to elastic rescattering.

\subsection{Chemical freeze-out properties}\label{maths}

To study flavor composition at chemical freeze-out we extract integrated particle yields extrapolating our measurement over the whole phases-pace.  In the framework of a chemical-equilibrium model~\cite{new2,Nu}, integrated yield ratios can be described by a set of parameters: the chemical freeze-out temperature ($T_{ch}$), the baryon and strangeness chemical potentials ($\mu_{B}$, $\mu_{s}$), and the strangeness suppression factor ($\gamma_{s}$). We fit our  ratios with such a model to extract these parameters. The value obtained for the chemical potential is independent of centrality within errors, and $\mu_{s}$ is consistent with 0. The obtained $\gamma_{s}$ increases quickly from peripheral to mid-central collisions approaching unity for most central events. In the framework of the model used this is an indication of  s-quark equilibration achieved in central heavy-ion collisions at RHIC.

In order to gain insights into collision evolution, we contrast $T_{ch}$ in Fig.~2  with the fitted kinetic freeze-out temperatures. Contrary to  $T_{kin}$, no centrality dependence is observed for $T_{ch}$. 
Moreover, the extracted $T_{ch}$ value is close to the QCD prediction of critical temperature. As hadronization is a universal process, it is suggestive that chemical freeze-out in heavy ion collisions at RHIC may  occur very close to hadronization~\cite{BM}.
In addition, multi-strange baryons appear to decouple earlier from the system, possibly at/right after chemical freeze-out. This  allows to estimate radial flow velocity at chemical freeze-out/hadronization  to be on the order of $0.45\pm 0.10$~$c$. Such a  significant amount of radial flow has to come from early partonic stage of the collision,  if indeed present at hadronization.

\begin{figure}[htb]
\insertplot{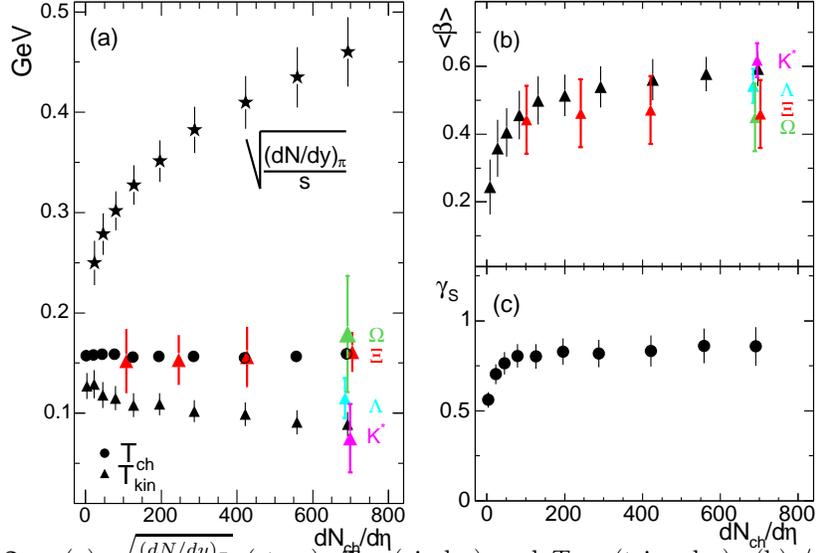}
\vspace*{-1.cm}
\caption[]{
(a) $\sqrt{\frac{(\dndy)_{\pi}}{S}}$
(stars), $T_{ch}$ 
(circles) and $T_{kin}$ 
(triangles); 
 (b) $\langle \beta \rangle$ and
(c) $\gamma_s$
as a function of the charged  hadron multiplicity. 
Black symbols represent results extracted from $\pi$, $K$, $p$ measurements and shown with systematic errors.
Preliminary results for rare particles are shown in colored symbols (as marked on the plot),  error bars are statistical only.
}
\label{fig2}
\end{figure}

\subsection{Inferred initial conditions}

To characterize initial condition we construct the following variable: $\frac{(dN/dy)_{\pi}}{S}$,
where $S$ is an estimate of the transverse overlap area  based on the number of participants, calculated by the MC Glauber model~\cite{pi_star}. The $\frac{(dN/dy)_{\pi}}{S}$  is related to the initial conditions of the collision such as energy density~\cite{Bjorken}. It is also a relevant quantity  in parton saturation picture, which suggests that in high energy collisions the initial gluon density is saturated up to a momentum scale that is proportional to $\sqrt{ \frac{ (dN/dy)_{\pi} }  {S} }$~\cite{zxu2}. 
The centrality dependence of $\sqrt{\frac{(dN/dy)_{\pi}}{S}}$ is shown in Figure 2, where one  sees a strong increase as function of collision centrality. In other words,  higher energy densities are achieved in more central collisions. 
However, despite of the significant change in the initial conditions, the collision  system yet always  evolve towards the same chemical freeze-out temperature.

Using measured  pion, kaon and (anti)proton spectra we estimate a mean transverse energy per particle in the 5\% most central events to be $\mEt \approx 600$~MeV. At RHIC energies the pion multiplicity in the final state correspond closely to the initial gluon multiplicity~\cite{KhN}. 
This would imply an initial temperature of $T \approx 300$~MeV assuming the ideal relativistic gas dynamics.  
For the case of 2 quark flavors, such temperature would give an initial energy density $\epsilon = \frac{37 \pi^2}{30}T^4\approx 10$~GeV~\cite{ref2}. This is in agreement with Bjorken energy density estimates by STAR~\cite{pi_star} and PHENIX~\cite{phenix}, and is significantly above the QCD predicted critical energy density,  supporting that a new state of matter created at RHIC in the early stage of the Au+Au collisions.

\section{Conclusions}\label{concl}

The Study of a variety of hadron spectra from STAR within the framework of chemical and local kinetic equilibrium model yields the following conclusions: $1)$~initial energy density increases with centrality and is sufficiently above the QCD predicted critical energy density. $2)$~Chemical freeze-out seems to occur at a universal temperature,  close to the predicted critical temperature for the phase transition. 
$3)$~Rare particles  appear to freeze-out kinetically with significant amount of flow at/right after chemical freeze-out.
$4)$~ The bulk of the medium evolves further towards final  kinetic  freeze-out with  lower temperature and larger flow velocity in more central collisions.

\vfill\eject
\end{document}